\documentclass[referee, sn-basic, Numbered]{sn-jnl}

\usepackage{graphicx}%
\usepackage{multirow}%
\usepackage{amsmath,amssymb,amsfonts}%
\usepackage{amsthm}%
\usepackage{mathrsfs}%
\usepackage[title]{appendix}%
\usepackage{xcolor}%
\usepackage{textcomp}%
\usepackage{manyfoot}%
\usepackage{booktabs}%
\usepackage{algorithm}%
\usepackage{algorithmicx}%
\usepackage{algpseudocode}%
\usepackage{listings}%
\usepackage{amsmath}
\usepackage{diagbox}
\usepackage{hyperref}
\usepackage{xpatch}
\usepackage{bm}
\usepackage{braket}  
\usepackage{geometry}%
\usepackage{ulem}%

\newcommand{\fref}{Fig.\ \ref}
\newcommand{\secref}{Section\ \ref}

\newcommand{\tref}{Table\ \ref}

\theoremstyle{thmstyleone}%
\theoremstyle{thmstyletwo}%

\theoremstyle{thmstylethree}%

\begin{document}
\let\WriteBookmarks\relax
\def\floatpagepagefraction{1}
\def\textpagefraction{.001}
\title[Article Title]{An Efficient Quantum Approximate Optimization Algorithm with Fixed Linear Ramp Schedule for Truss Structure Optimization}

\author[1]{\fnm{Junsen} \sur{Xiao}}\email{xiao.junsen.s2@dc.tohoku.ac.jp}

\author[2]{\fnm{Naruethep} \sur{Sukulthanasorn}}\email{sukulthanasorn.naruethep.c2@tohoku.ac.jp }

\author[2]{\fnm{Reika} \sur{Nomura}}\email{rnomura@tohoku.ac.jp}

\author[2]{\fnm{Shuji} \sur{Moriguchi}}\email{shuji.moriguchi.d6@tohoku.ac.jp}

\author*[1]{\fnm{Kenjiro} \sur{Terada}}\email{tei@tohoku.ac.jp}
\affil*[1]{\orgdiv{Department of Civil and Environmental Engineering}, \orgname{Tohoku University}, \orgaddress{\street{Aza-Aoba, 468-1, Aramaki, Aoba-ku}, \city{Sendai}, \postcode{980-8572}, \country{Japan}}}

\affil[2]{\orgdiv{International Research Institute of Disaster Science}, \orgname{Tohoku University}, \orgaddress{\street{Aza-Aoba, 468-1, Aramaki, Aoba-ku}, \city{Sendai}, \postcode{980-8572}, \country{Japan}}}

  
 
 

\abstract{
This study proposes a novel structural optimization framework based on quantum variational circuits, in which the multiplier acting on the cross-sectional area of each rod in a truss structure as an updater is used as a design variable.
Specifically, we employ a classical processor for structural analysis with the finite element method, and the Quantum Approximate Optimization Algorithm (QAOA) is subsequently performed to update the cross-sectional area so that the compliance is minimized. 
The advantages of this framework can be seen in three key aspects. First, by defining design variables as multipliers, rather than simply reducing the design variable to a binary candidate of inclusion or exclusion (corresponding to qubit states, ``0" and ``1"), it provides greater flexibility in adjusting the cross-sectional area of the rod at each iteration of the optimization process.
Second, the multipliers acting on rods are encoded with on-off encoding, eliminating additional constraints in the convergence judgement. As a result, the objective function is in a simple format, enabling efficient optimization using QAOA.
Third, a fixed linear ramp schedule (FLRS) for variational parameter setting bypasses the classical optimization process, thereby improving the operational efficiency of the framework.
In the two structural cases investigated in this study, the proposed approach highlights the feasibility and applicability potential of quantum computing in advancing engineering design and optimization. 
Numerical experiments have demonstrated the effectiveness of this framework, providing a firm foundation for future research on quantum-assisted optimization methods in engineering fields.
}




\maketitle
\section{Introduction}\label{sec1}
Structural optimization is a computational design method aimed at determining the optimal size of structural members, geometrical shape, or distribution of materials to enhance performance while adhering to given constraints, such as material volume, stress distribution, and target strength. Since the pioneering works by Bendsøe and Kikuchi \cite{bendsoe2013topology, bendsoe1988generating}, topology optimization has evolved in particular and has come to be applied to various fields \cite{kiziltas2003topology, nomura2007structural, FUJII2020120082, IRADUKUNDA2020115723, CHEN2020112806, CHEN2021107054, yesilyurt2021efficient, mukherjee2021accelerating}. So far, numerous optimization algorithms have been developed to achieve efficient and effective structural designs. Prominent techniques include gradient-based methods \cite{Raphael2012elements, Andreassen2011matlab}, genetic algorithms \cite{DEB2001447, WANG20053749}, and evolutionary structural optimization \cite{hu2020fracture}, complemented by advanced computational strategies such as parallel computing \cite{MAKSUM2022100352}, dimensionality reduction strategies \cite{Gogu2015reduce}, and geometric primitives \cite{GUO2022103238}. While these approaches have historically demonstrated considerable success, recent years have witnessed challenges in achieving significant improvements in computational performance, largely due to the saturation of algorithmic advancements and the stagnation of progress in classical computational hardware \cite{moore1998cramming, Xu2024Gate}. To break through these bottlenecks, innovative optimization frameworks and technologies capable of surpassing current limitations and driving substantial advancements in computational efficiency are further required.

In recent years, quantum computing (QC) has gained tremendous attraction as an emerging and alternative computing paradigm that can theoretically be superior to classical computing for a variety of optimization problems\cite{Frank2019quantum}, including structural optimization. This superiority stems from the entanglement and superposition of quantum states. Generally, candidate solutions are encoded through a string of qubits in QC. The qubits can exist in 0 and 1 simultaneously, while the classical bit must be either 0 or 1. This superposition allows QC to search multiple solutions in parallel. With the theoretical foundation, QC has been applied to optimization problems in various disciplines in recent years. Brandhofe et al. \cite{brandhofer2022benchmarking} conducted a benchmarking test of portfolio optimization problems utilizing quantum optimization algorithms\cite{brandhofer2022benchmarking}. Xiao et al. \cite{xiao2024fmqa} examined the applicability of quantum computing for hyperparameter optimization and metamodel-based granular flow simulation optimization. Kurowski et al.\cite{KUROWSKI2023518} applied a quantum variational circuit to the job shop scheduling problem. 

To date, there are currently two main types of quantum hardware in the current stage as the noisy intermediate-scale quantum (NISQ) devices: gate-based universal quantum computers (GQC)\cite{Deutsch1985universal} and quantum annealers (QA) embrace the annealing techniques\cite{kadowaki1998annealing}. The former alters the amplitudes of the states in the superimposed Hamiltonian through a series of unitary operators (quantum gates), enabling the ground state to be observed with sufficiently high probability.
The latter obtains the ground state of the Hamiltonian based on the adiabatic evolution\cite{born1928beweis} of quantum fluctuations. Currently, the number of qubits that can be adopted in QA is roughly in the thousands, but only a few dozen can be reliably used in GQC. In addition, reliably executable GQC circuits are necessarily constrained to remain short in depth, because of the inherent noise in quantum gates and the substantial overhead associated with error mitigation. Given these circumstances, QA has been successfully applied to structural optimization problems with different proposed approaches \cite{honda2024development,ye2023quantum,sukulthanasorn2024quantum}. Nevertheless, since QA hardware is specifically designed for the application as optimization through the annealing process, its applicability across a broad range of applications may be limited and could potentially be replaced by the GQC in the long term. 

Based on existing literature, one of the most well-known algorithms tailored for solving optimization problems by GQC is Quantum Approximate Optimization Algorithm (QAOA)\cite{Edward2014qaoa}. It prepares the problem's Hamiltonian by a variational quantum circuit composed of a series of quantum operators, and the variational parameters in the operators are finely tuned by adiabatic evolution (consistent with QA) and a layering strategy. The circuit consists of phase-separating and mixer Hamiltonian operators, which are alternately applied, with each pair of operators corresponding to a layer. If the number of layers is denoted by $p$, then when $p \to \infty$, the circuit is theoretically closer to an ideal adiabatic evolution. 
The phase-separating operators symbolize the problem' Hamiltonian as a combination of multiple quantum gates, whereas the mixer operators alter the amplitudes of the states in Hamiltonian. The basis state can be observed with a sufficiently high probability once the variational parameters are optimized by a classical processor. In recent years, the application of QAOA has gradually expanded from graph theory problems such as Max-Cut\cite{Wang2018max}, multi-knapsack\cite{Awasthi2023}, and Max-Sat\cite{Yu2023sat}, to real-world optimization problems\cite{Pontus2020tail,Wang2012insp,Babbush2018depth}. Along with this trend, some researches have also proceeded to design the mixer operators to improve the approximation accuracy of QAOA\cite{marsh2019quantum,Wang2020xy,kim2023quantum}. So far, the application of QAOA to structural optimization and other computational mechanics problems remains limited, highlighting the need to explore a suitable integration framework between QAOA and structural optimization.

To the best of authors' knowledge, Kim and Wang\cite{Kim2023to} first applied QAOA to truss optimization problems. They directly defined the existing rod as binary values, representing the addition or removal of rods in a truss structure. All possible candidate rod members are incorporated into the objective function, formulating the problem as a well-known ground structure optimization problem. Although they successfully obtained a solution with acceptable accuracy for a small-scale truss structure with 6 rods, this approach faces challenges when extending to larger problem sizes due to the computational expense of encoding candidate solutions into the objective function. Consequently, the complexity of the objective function increases exponentially with the number of rods, resulting in significant computational time requirements. Additionally, Xu et al.\cite{Xu2024Gate} presented the alternative approach for representing the cross-sectional areas of rods as discrete variables using one-hot encoding, and defined an objective function that minimizes the volume of material usage to meet the load and deformation requirements. Their approach ensures that the objective function remains quadratic, but the cost of using one-hot encoding is a several-fold increase in qubits consumption. Besides, it should be noted here that the QAOA circuit parameters also require to be fine-tuned with iterative optimization. This makes structural optimization using QAOA expensive and remains a challenge in enhancing computational efficiency. 

In this study, taking the structural optimization problem of truss structures as an example, we propose an efficient iterative updating framework that effectively uses an updater that adjusts the cross-sectional area based on the optimization results obtained from QAOA. 
Furthermore, by using predetermined parameters in the quantum variational circuit, this framework eliminates the need for iterative fine-tuning by a classical processor, and we have successfully applied this approach to the truss optimization problem for the first time.
To be more specific, in order to minimize compliance while meeting a volume constraint, the cross-sectional areas and volumes are adjusted in each design iteration by multiplying the cross-sectional area of each rod by the updater obtained from QAOA.
In the present study, 
each updater is encoded by on-off encoding using two qubits, and the objective function is expressed as a quadratic function. 
This setting allows the cross-sectional area of each rod to be dynamically and flexibly adjusted during the optimization process.
Specifically, the cross-sectional area of a rod that has experienced high strain energy gradually increases, while the cross-sectional area of a rod that has experienced low strain energy decreases. 
As a result, the arrangement of truss members (rods) gradually converges to an optimal configuration.
In order to prevent hardware noise, this research uses a quantum simulator rather than an actual quantum processor. 
This simulator is provided by \texttt{Pennylane 0.39}, an open-source Python library for quantum programming.

\section{Quantum Approximate Optimization Algorithm}\label{sec2}
\begin{figure}[htbp]
    \centering
    \includegraphics[width=0.6\linewidth]{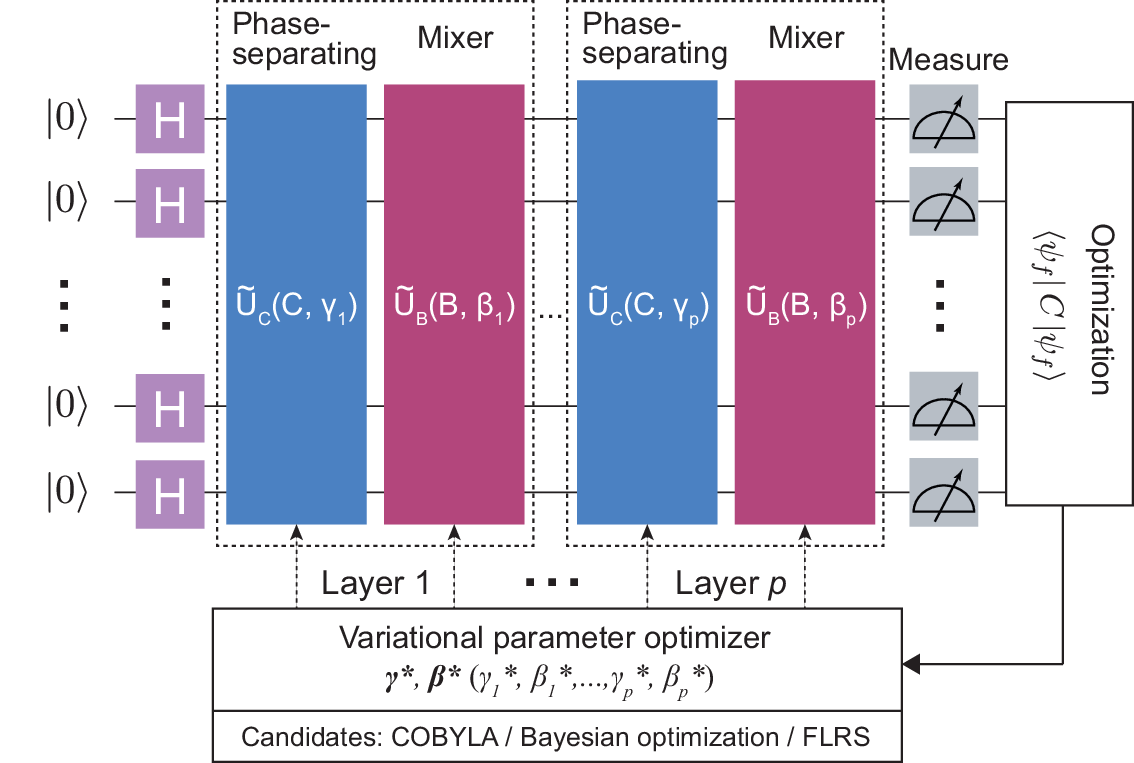}
    \caption{Architecture of QAOA circuit.
}
\label{fig:circuit}
\end{figure}

In this section, we briefly summarize the architecture of the QAOA circuit and discuss fine-tuning approaches for QAOA circuit parameters, called variational parameters.
\subsection{Combinatorial optimization problem and QAOA principle}
Many interesting real-world problems can be framed as
combinatorial optimization problems, which are aimed at finding an optimal object from a finite set of objects. A combinatorial optimization problem can be phrased as a maximization of an objective function which is a sum of Boolean functions. Each Boolean function $C_d(\bm{z}) : \{-1, 1\}^{n} \rightarrow \{-1, 1\}$ takes the qubit string $z = z_1z_2\cdots z_n$ as input and a Boolean value (-1 or 1) as output. The objective function in Hamiltonian form can then be formulated as
\begin{equation}
    C = \sum_{d=1}^{D}{w_dC_d(\bm{z})}\label{eq1_comb}{,}
\end{equation}
where $D$ is the number of Boolean functions. In more general problems, $w_d$ are added as factors to introduce weights to different objects. QAOA was originally designed to leverage quantum acceleration for solving such combinatorial optimization problems by Farhi et al.\cite{Edward2014qaoa}. They took inspiration from the Trotterized version of the quantum adiabatic evolution theorem \cite{born1928beweis}. The evolution of a $n$-qubits system performed by QAOA quantum circuit is defined as
\begin{equation}
    \ket{\psi} = \tilde{U}_{B}(B, \beta_p)\tilde{U}_{C}(C, \gamma_p) \cdots \tilde{U}_{B}(B, \beta_1)\tilde{U}_{C}(C, \gamma_1)\ket{+}^{\otimes n}\label{eq2_qaoa1}{.}
\end{equation}
Here, $\ket{+}^{\otimes n}$ is uniform superposition initial state for the $n$-qubits system, which prepared by applying Hadamard gates to all qubits with the initial state at $\ket{0}$. Also, $\tilde{U}_{C}$ and $\tilde{U}_{B}$ represent the phase-separating operators and mixer operators, respectively. These two operators are alternately mounted on the circuit.

The phase-separating operator $\tilde{U}_{C}(C, \gamma)$ symbolizes the problem's Hamiltonian $C$ with multiple quantum gates. In particular, $C$ is obtained from the conventional objective function with Ising variables $z \in \{-1, +1\}$, which can be easily transferred from conventional binary variables $q \in \{0, 1\}$ as 
\begin{equation}
    q_j \rightarrow \frac{1}{2}(1-z_j) \quad (j = 1, \ldots ,n)\label{eq3_bintois}{.}
\end{equation}
The Ising variables are encoded in $\tilde{U}_{C}$, which applies a Pauli $Z$ matrix on each qubit. Specifically, given the problem's Hamiltonian $C$ and variational parameter $\gamma$, $\tilde{U}_{C}$ is defined as 
\begin{equation}
   \tilde{U}_{C}(C, \gamma) = e^{-i\gamma C} = \prod_{d=1}^{D}e^{-i\gamma w_dC_d(Z)}\label{eq4_uc}{.}
\end{equation}
On the other hand, the mixer operator $\tilde{U}_{B}$ alters the amplitudes of the states in Hamiltonian; i.e., a Pauli $X$ gate $\sigma^{x}$ is applied to each qubit in the circuit. The Pauli $X$ gate gives each qubit a counterclockwise rotation around the Bloch sphere along the $X$ axis, represented by the variational parameter $\beta$. To be specific, given the variational parameter $\beta$, $\tilde{U}_{B}$ is defined as 
\begin{equation}
   \tilde{U}_{B}(B, \beta) = e^{-i\beta B} = \prod_{j=1}^{n}e^{-i\beta \sigma^{x}_{j}},\text{ where } B = \sum_{j=1}^{n}{\sigma^{x}_{j}}\label{eq5_uB}{.}
\end{equation}
After repeating the sequence of $\tilde{U}_{C}$ and $\tilde{U}_{B}$ $p$ times, the measurement operators converges $\ket{\psi}$ to the single basis state $\ket{\psi_f}$, which is the optimal solution identified by the algorithm. 
The objective function value corresponding to the expectation of $C$ is approximately computed as  
\begin{equation}
   f(C, \ket{\psi_f}) \approx \bra{\psi_f} C \ket{\psi_f} \label{eq6_exp}{.}
\end{equation}

QAOA is a quantum-classical hybrid approach, with some of the classical processor acting as an optimizer for the variational parameters in the circuit as shown in \fref{fig:circuit}. The goal of the above-described process is to find the optimal set of variational parameters $(\bm{\gamma}^{*}, \bm{\beta}^{*})$ such that the expectation value $f(C, \ket{\psi_f})$ is maximized as

\begin{equation}
   (\bm{\gamma}^{*}, \bm{\beta}^{*}) = \underbrace{\mathrm{arg} \; \mathrm{max}}_{\forall \; \bm{\gamma}, \bm{\beta}}f(C, \ket{\psi_f}) \label{eq7_para}{.}
\end{equation}

Typically, the performance of the QAOA greatly depends on the classical optimizer adopted for determining the variational parameters. When the phase-separating and mixer Hamiltonian operators are alternately applied $p$ times, the circuit depth QAOA corresponds to $p$ layers, necessitating the optimization to determine $2p$ variational parameters. 
As mentioned in Introduction, as $p \rightarrow \infty$, the QAOA circuit approaches ideal adiabatic evolution and achieves optimal precision in theory\cite{Shor2014adia}.

\subsection{Fine-tuning of QAOA circuit parameters}
The classical optimizers for fine-tuning QAOA circuit parameters that are widely recognized today include gradient-based Broyden-Fletcher-Goldfarb-Shanno (BFGS) \cite{Flecher1992} algorithms and linearly approximation approach COBYLA. 
Additionally, Kim and Wang \cite{kim2023quantum} proposed QABOA, which adopted Bayesian optimization as a classical optimizer. However, even with moderate $p$-values (e.g., 5–10), they face the barren plateau problem\cite{wang2021noise}, making it difficult to achieve global optimum. Furthermore, in the QAOA circuit, fine-tuning of its parameters is required before designing the target structure, and this can be another iterative process,  which is independent of iterative structural updating. 
This double iteration significantly reduces computational efficiency, making the QAOA approach less practical. Recent investigations have focused on this issue, and some studies have explored patterns of variational parameters to simplify the optimization process. 
For instance, Zhou et al.\cite{Zhou2020devices} and Montanez-Barrer and Michielsen\cite{Montanez2024protocol} identified general trends in optimal variational parameters in large-scale Max-cut problems. 
The optimal parameters, $\gamma^{*}$ and $\beta^{*}$, increase and decreases steadily with $p$, respectively. 
The former study proposed a strategy that optimizes the initial variational parameters ($p = 1$) by BFGS, then iteratively sets the subsequent variational parameters ($p > 1$) using linear interpolation. 
The latter suggested using a fixed linear ramp schedule (FLRS) for variational parameter settings. Both approaches bypass most of the optimization process and achieve promising results.

In this study, the FLRS method is combined with the proposed update design scheme to significantly reduce
the computational cost associated with fine-tuning the QAOA circuit parameters. 
To investigate the applicability of QAOA to truss optimization problems, we conduct a detailed evaluation of the performance comparison of COBYLA, Bayesian optimization, and FLRS in terms of computational cost and accuracy for two optimization cases with $p$ set to 6 and 8. 

As a point to note, there are two unavoidable issues in the practical implementation of QAOA. 
First, it is not possible to directly access the observable expectation at each location in the QAOA circuit, and since measurements are limited to one per circuit execution, it is necessary to estimate the necessary information by executing the circuit many times to obtain many measurements of the problem variable and averaging them. 
Thus, the expectation can only be approximated by the average, and its uncertainty is inversely proportional to the square root of the number of measurements. In this study, the number of measurements is set to $10^{5}$ to mitigate the inherent uncertainty as much as possible. 
The second problem is that 
the result reported at the end of the QAOA experiment is the most frequently measured qubit string among all candidates, not the average value (expectation) of $C$ \cite{kim2023quantum,Larkin2022QAOA}. 

\section{QAOA-based framework for truss optimization}\label{sec3}

In this section, the details of the QAOA-based structural optimization framework is outlined 
using two examples of truss optimization.
After setting the static equilibrium problem of truss structures, we explain the iterative updating scheme, the encoding operation on design variables, and the definition of the objective function in turn.

\begin{figure}[htbp]
    \centering
    \includegraphics[width=0.6\linewidth]{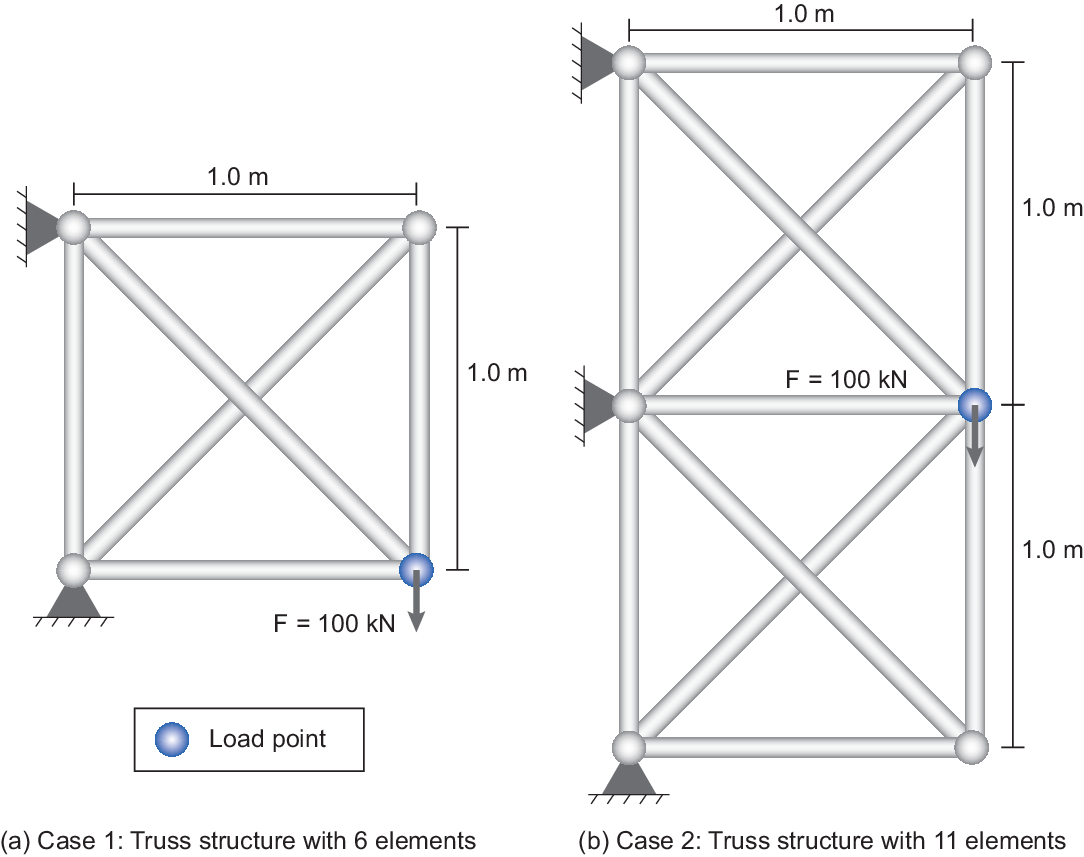}
    \caption{2D truss structures.
}
\label{fig:truss}
\end{figure}

\subsection{Truss structure setup}\label{subsec31}
This study configures two 2D truss structures consisting of 6 and 11 rods, as shown in \fref{fig:truss}, as optimization targets. Each rod has a length $L$ of either 1.0 m or $\sqrt{2}$ m, with a Young's modulus $E$ of $2\times10^{11}$ $\mathrm{N/m^2}$, and a initial cross-sectional area $A^{(0)}$ of 0.5 $\mathrm{m^2}$. Then, the initial volumes $V^{(0)}$ for Case 1 and Case 2 can be calculated as 3.414 $\mathrm{m^2}$ and 6.328 $\mathrm{m^2}$, respectively, which serve as volume constraints associated with the objective function for QAOA. A load of 100 kN is applied on the blue node, acting vertically downward. The vertical rods on the left in both of these cases do not transmit any load, but are still included in the optimization target for verification purposes. That is, with this inclusion, we can verify that QAOA is properly working by checking whether these rods are removed in the first few iterations. Note that the self-weight of the structure is not considered in this study.

\begin{figure}[htbp]
    \centering
    \includegraphics[width=0.75\linewidth]{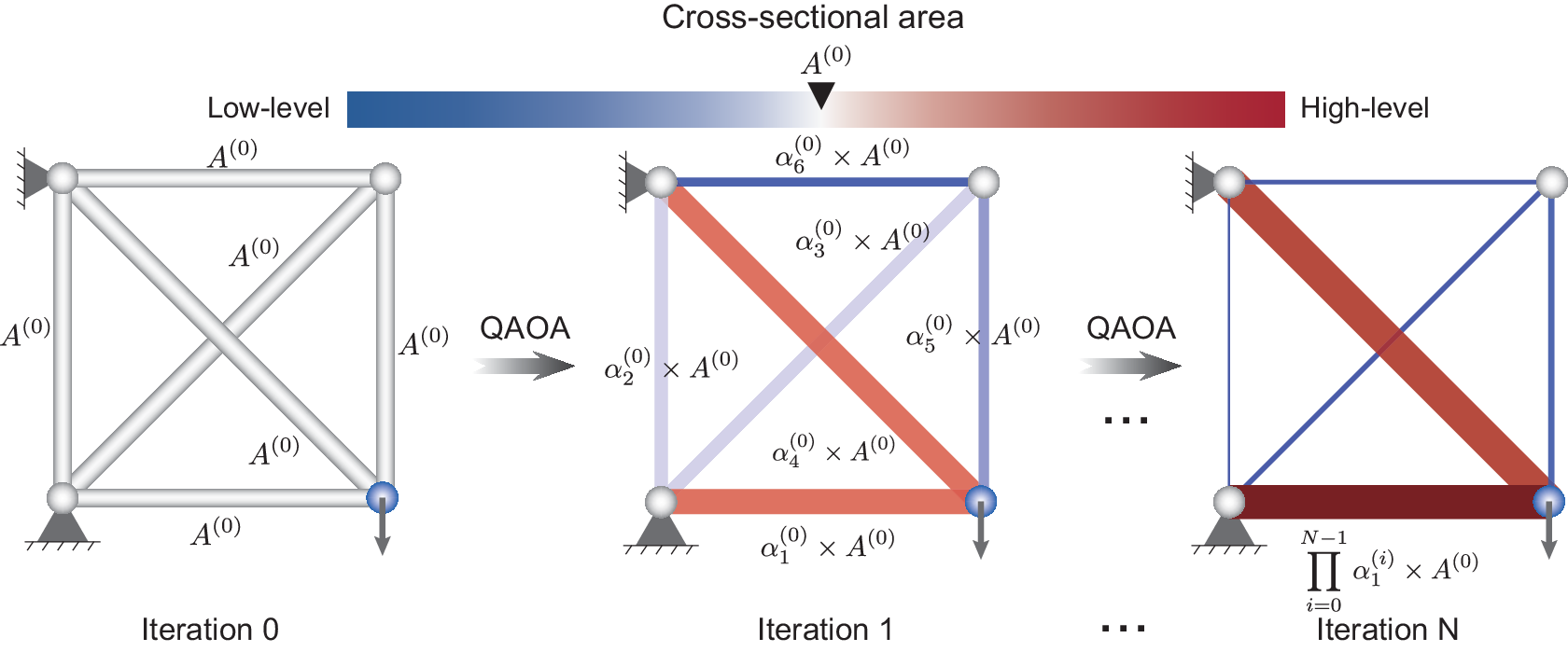}
    \caption{Iterative updating scheme.
}
\label{fig:update}
\end{figure}

\subsection{Iterative updating scheme and design variables}\label{subsec32}
As mentioned in Introduction, we design an optimization framework characterized by iterative updates, in which the multiplier $\alpha$ acting on cross-sectional areas $A$ is defined as a design variable.  
The purpose of iterative updates is to minimize compliance while satisfying a volume constraint, and the multiplier works to adjust the cross-sectional area of each rod in each iteration.  
Taking Case 1 as an example, we can envision the optimization problem encountered by QAOA in each iteration with \fref{fig:update}.

According to the initial setup of the truss structure, before the iteration starts (Iteration 0 in \fref{fig:update}), we assume the same cross sectional area $A^{(0)}$ for all rods.
Subsequently, the cross-sectional areas is updated by multiplying by a series of $\alpha$, and for this reason, this multiplier is referred to below as ``updater''.
In the figure, the superscript on $\alpha$ and $A$ represents the iteration index $i$, which can be assigned values of (0, 1, \ldots, $N-1$), and the subscript on $\alpha$ represents the rod number $e$, which takes values of ($e$ = 1, 2, 3, 4, 5, 6) in Case 1. 
Since the $\alpha$ is the design variable, QAOA is responsible for determining the value of the updater for each rod in each iteration based on the objective function. The range of updater is specified as 
\begin{equation}
   0 < \alpha \leq \theta \label{eq8_upda}{,}
\end{equation}
where the $\theta$ acts as an upper limit that the updaters can be set to within each iteration. In this framework, this value can be customized in a specific problem. It is obvious that when $\alpha > 1.0$, the cross-sectional area of the corresponding rod will be increased in this iterative process, and vice versa. For example, if $\alpha_1^{(0)}$ is optimized to 1.1, the area of rod 1 will be $1.1 \times A^{(0)}$; if $\alpha_2^{(0)}$ is optimized to 0.5, the area of rod 2 will be $0.5 \times A^{(0)}$. 
This operation efficiently allocates materials to each rod of the truss structure to meet volume constraints.
After $N$ iterations, the cross-sectional area of rod $e$ is updated such that
\begin{align}
   A_e^{N-1} &= \alpha_e^{(N-1)} \cdot\alpha_e^{(N-2)} \cdots \alpha_e^{(2)} \cdot\alpha_e^{(1)} \cdot\alpha_e^{(0)} \cdot A^{(0)}\label{eq9_updato}
   \\ 
   &= \prod_{i=0}^{N-1}{\alpha_e^{(i)} A^{(0)}} \label{eq10_updato}{.}
\end{align}
This update will ultimately be reflected in the stiffness matrix of this rod as
\begin{align} 
   \bm{K}_{e}^{(N-1)}= \prod_{i=0}^{N-1}{\alpha_e^{(i)} \bm{K}_{e}^{(0)}} =  \prod_{i=0}^{N-1} {\alpha_e^{(i)}} \frac{E{A}^{(0)}}{L_e} \bm{J}
\label{eq11_updak}{,}
\end{align}
where $L_e$ is the length of rod $e$, and $\bm{J}$ is a square matrix of order 4 whose constant components are determined by the orientation of the rod axis. 

It should be noted that the conventional method for structural optimization such as the optimality criteria (OC) method\cite{Andreassen2011matlab} can easily define the updater $\alpha$ as a continuous variable. However, in the current situation, this cannot be easily implemented on a quantum processor for both QA and GQC. 
In fact, the generally accepted method requires an encoding operation that uses multiple binary variables (qubits) as discrete variables.
Ideally, a discrete variable encoded by a sufficient number of qubits could be rich enough to an approximate continuous variable. 

So far, several encoding methods have been proposed, including one-hot encoding, binary encoding, on-off encoding\cite{endo2024encod}, etc. 
Xu et al.\cite{Xu2024Gate} proposed a non-iterative framework based on QAOA and implemented an algorithm for truss optimization with one-hot encoding. However, their method differs from ours in that they directly set the cross-sectional area of each rod as a design variable. 
The area of a rod $e$ is then encoded by multiplying a series of qubits using a pre-determined corresponding candidate vector $\bm{r}$
as follows: 
\begin{equation}
       A_{e} = \sum_{m=1}^{M}{r_{m}q_{e, m}} \text{ }(q \in \{0, 1\})\quad \textrm{s.t.} \quad \sum_{m=1}^{M}{q_{e, m}} =1
       \label{eq12_onehotencod}{,}
\end{equation}
where $M$ represents the number of pre-determined candidate values, $q$ refers to the qubits (binary variables) adopted for the rod $e$. Because one-hot encoding strictly enforces that for a given rod $e$, only one qubit in the associated set can be in state ``1". Therefore, it lacks a combinatorial effect, and $m$ candidate values should be expressed using $m$ qubits.
In their study, since five candidates were assigned to each rod for a simple truss structure consisting of three rods, fifteen qubits were required in total. 
On the other hand, Sukulthanasorn et al.\cite{sukulthanasorn2024quantum} proposed a framework for truss and topology optimization using quantum annealing, in which only one qubit per rod or element is used to control the updater value. 
In the structural optimization of trusses using this method, the state ``1" represents an increase in the cross-sectional area of the rod, while the state ``0" signifies a decrease. This approach saves qubits, but when the truss volume approaches the upper limit of the volume constraint, there are not enough candidates to maintain a constant cross-sectional area. 
When the truss volume approaches the volume constraint, the design iteration is stopped manually by an appropriate criterion. 

In the framework proposed in this study, on-off encoding is adopted to encode the updater. 
In on-off encoding, the update data can be encoded as a linear sum of predefined digits, and the ``0" and ``1" states of each qubit can be controlled without constraint.
Specifically, for a certain rod $e$, the updater $\alpha_e$ is encoded as
\begin{equation}
       \alpha_{e} = \sum_{m=1}^{M}{r_{m}q_{e, m}} \text{ }(q \in \{0, 1\})
       \label{eq13_onoff}{.}
\end{equation}
It is clear that on-off encoding has a combinatorial effect,
and $m$ qubits can encode $2^m$ candidate values. 
For example, if the predefined candidate vector is $\{0.1, 1.0\}$ in on-off encoding using two qubits, the updater with fousr options is encoded as $\{0.0, 0.1, 1.0, 1.1\}$.
In this study, as in the case of this example, two qubits are assigned to update the cross-sectional area of one rod in the truss structure. If one of the options is set to 1.0, QAOA does not change the cross-sectional area, so the truss structure automatically and stably converges as the truss volume approaches the volume constraint.

It should be noted that as the number of qubits $m$ increases, the updater can be encoded with richer expressive performance, 
and it becomes possible to express it close to a continuous variable. 
However, this comes at the cost of an exponential expansion (to the power of 2$m$, i.e., $2^m$) in the number of search combinations.
The reason for encoding each updater with two qubits in this study is that we considered the
trade-off between automatic convergence and qubit consumption. 
Fortunately, in our framework, the cross-sectional area of a rod is updated iteratively by multiplying the updater $\alpha$, so the actual representation of the cross-sectional area itself gradually becomes richer with each iteration.
In fact, the research results of Sukulthanasorn et al.\cite{sukulthanasorn2024quantum} showed that 
even with only two qubits, the change in the cross-sectional area of a rod achieved sufficient flexibility. 
At the very least, it can be said that it is not absolutely necessary to allocate additional qubits to a single rod. 
Therefore, we believe that the framework we propose can be applied to large-scale structural optimization problems and that it can also save qubits. 
In the next section, we will introduce the definition of the QAOA objective function in detail.

\subsection{Objective function for QAOA}\label{subsec33}
As mentioned in \secref{subsec32}, the purpose of iterative updating is to minimize the compliance of a truss structure while satisfying a volume constraint.
Assuming that the target truss structure has a total of $N_{e}$ rods, the optimization problem that maximizes mean compliance can be formulated in the following standard format:
\begin{equation}
    \left\{
    \begin{array}{ll}
\text{find : } &\alpha \in \{ \alpha_1, \alpha_2, \ldots, \alpha_e, \ldots, \alpha_{N_e}\}\\[2ex]
\underset{\alpha_e}{\text{min : }} & \bm{F}^{T}\bm{U} 
 \text{ (compliance) }\\[2ex]
\text{s.t. : } &\bm{K}(\alpha)\bm{U} = \bm{F}\\[1ex]
   &\sum_{e=1}^{N_{e}}{V_e(\alpha_e)} \leq V^{(0)}\\[1ex]
   & 0 < \alpha_e \leq \theta\label{eq14_objective}
\end{array}{,}
    \right.
\end{equation}
where $\bm{F}$ is the external applied load vector, $\bm{U}$ is the global nodal displacement vector in matrix structural analysis, and $\bm{K}(\alpha)$ is the global stiffness matrix that depends on the updater $\alpha$ as in Eq. \eqref{eq11_updak}. 
Also, $\sum_{e=1}^{N_{e}}{V_e(\alpha_e)}$ refers to the volume of the entire truss structure in a specific iteration and is constrained so that it does not exceed the initial volume $V^{(0)}$. 
This is exactly the volume constraint in our framework. Usually, a slack variable $S$ is introduced to incorporate the volume constraint into the objective function. 
Specifically, the original inequality constraints can be rewritten as the following equality constraint by the slack variable: 
\begin{equation}
f_\textrm{cons} := \frac{\sum_{e=1}^{N_{e}}{V_e(\alpha_e)}}{V^{(0)}} + S -1 = 0\label{eq16_slack}{.}
\end{equation}
Additionally, since minimizing compliance is equivalent to maximizing structural stiffness, 
the objective function in Eq. \eqref{eq14_objective} can be rewritten as 
\begin{align}
\bm{F}^\top \bm{U} \bm{K}(\alpha) & =\bm{U}^\top \bm{K}(\alpha) \bm{U} 
=
\sum_{e=1}^{N_e}{ \bm{U}_e^\top \frac{A_e(\alpha_e) E}{L_e} \bm{J} \bm{U}_{e}} {,} 
\end{align}
where $\bm{U}_e$ denotes the element displacement vector in matrix structural analysis conducted on a classical processor. 

Using the above expressions for equality constraint and stiffness-based objective function, we rewrite the optimization problem in Eq. \eqref{eq14_objective} as
\begin{equation}
    \left\{
    \begin{array}{ll}
\text{find : } &\alpha \in \{ \alpha_1, \alpha_2, \ldots, \alpha_e, \ldots, \alpha_{N_e}, S\}\\[2ex]
\underset{\alpha_e}{\text{min : }} & 
 -\bm{U}^{T}\bm{K}(\alpha)\bm{U} + \lambda\left(\dfrac{\sum_{e=1}^{N_{e}}{V_e(\alpha_e)}}{V^{(0)}} + S -1\right)^2 := f_\textrm{obj}\\[2ex]
\text{s.t. : } &\bm{K}(\alpha)\bm{U} = \bm{F}, \text{ } 0 < \alpha_e \leq \theta\label{eq17_ob2}
\end{array} {.}
    \right.
\end{equation}
Here, the negative sign preceding the first term of the objective function arises from the reformulation of the minimization into a maximization problem, aiming to identify the updater that maximizes stiffness. Also, $\lambda$ is the weight coefficient for the volume constraint defined according to the problem scenario.
The definition of $\theta$ remains the same as in Eq. \eqref{eq8_upda}, serving as the upper limit that that can be set within each iteration.

As described in the previous section, each updater is encoded using on-off encoding with two qubits as in Eq. \eqref{eq13_onoff}, and the pre-determined candidate vector $\bm{r}$ is defined as $\{0.1, 1\}$. 
Thus, totally four options can be generated for each updater as $\alpha_e \in \{0.0, 0.1, 1.0, 1.1\}$. 
According to Eq. \eqref{eq8_upda} and Eq. \eqref{eq13_onoff}, $\theta$ is typically set to 1.1. 
It should be noted that when QAOA optimizes the value of a specific updater to 0.0, the proposed method replaces this zero with a random number $\epsilon$ in the range $[1 \times 10^{-10}, 2\times 10^{-10}]$ to prevent the stiffness matrix from becoming singular.
Note that the slack variable likewise needs to be encoded as a discrete variable by multiple binary variables through on-off encoding. In this study, this is realized by another set of qubits $q_c$ as
\begin{equation}
S(q_c) = \frac{\sum^{N_c}_{c=1}{k_cq_c}}{\sum^{N_c}_{c=1}{k_c}}\label{eq18_slackencod}{.}
\end{equation}
Here, $N_c$ is the total number of qubits used to encode the slack variable, $k_c \;(c=1, \ldots, N_c)$ are the coefficients constructing the candidate vector $\bm{k}$ pre-determined to encode the slack variable. Generally, the components in $\bm{k}$ are often defined as the exponential form of 2, such as 2, $2^{2},\ldots, 2^{N_c}$. In this study, to reduce the total qubit consumption, only two qubits are used to encode the slack variable, i.e., $N_c = 2$ and $\bm{k} = \left\{2, 2^2 \right\}$. The denominator in Eq. \eqref{eq18_slackencod} is used to normalize the slack variable to the range [0, 1]. 

By using Eq. \eqref{eq13_onoff} and Eq. \eqref{eq18_slackencod} in Eq. \eqref{eq17_ob2}, the objective function with binary variables, $q_e$ and $q_c$, can be expressed as
\begin{align}
   f_\textrm{obj}(q_e, q_c) &= f (q_e) + \lambda f_\textrm{cons}(q_c) \nonumber \\
    & = -\sum_{e=1}^{N_e}{ \bm{U}_e^\top \frac{ \sum_{m=1}^{M} r_{m} q_{e, m} \hat{A}_e E}{L_e} \bm{J} \bm{U}_{e}} + \lambda \left(\frac{\sum_{e=1}^{N_{e}}{V_e(q_e)}}{V^{(0)}} +\frac{\sum^{N_c}_{c=1}{k_cq_c}}{\sum^{N_c}_{c=1}{k_c}} -1 \right)^2{,}
    \label{eq20_obj3}
\end{align}
where $\hat{A}_e$ is an updated cross-sectional area of rod $e$ in the previous iteration.
If the problem Hamiltonian $C$ in Eq. \eqref{eq1_comb} is constructed in the form of Eq. \eqref{eq20_obj3}, a quantum processor can be employed to solve this optimization problem. This step corresponds to the grey arrow between iterations in \fref{fig:update}. 
To provide a more intuitive and clear understanding, a detailed process and flowchart will be provided in the next section.

\begin{figure}[htbp]
    \centering
    \includegraphics[width=0.6\linewidth]{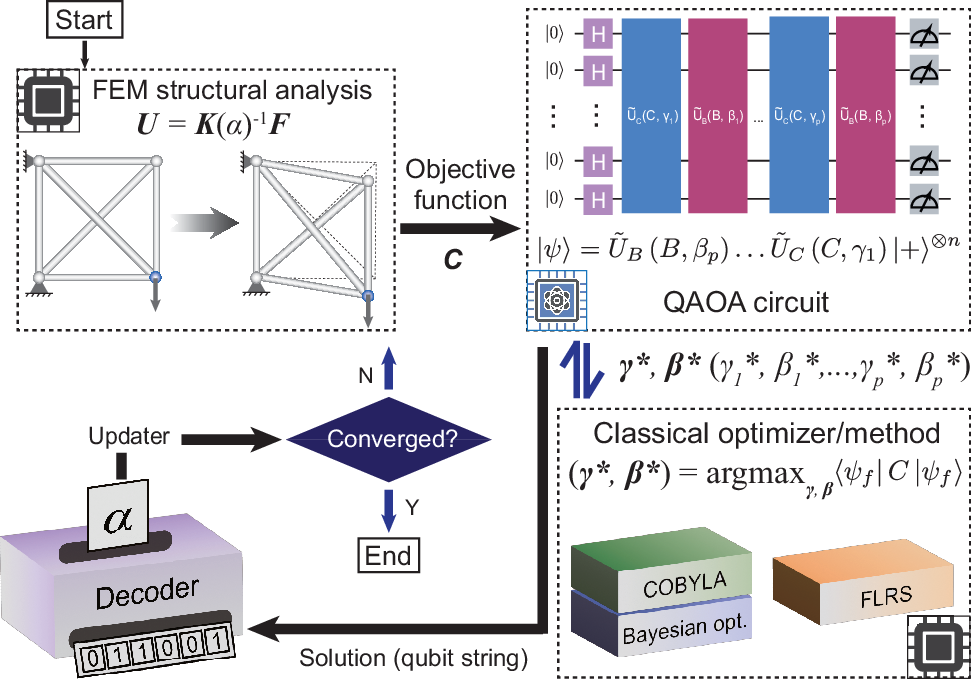}
    \caption{Flowchart for the proposed framework.
}
\label{fig:flowchart}
\end{figure}

\subsection{Overall process of proposed framework}\label{subsec34}
To briefly describe the proposed framework, the specific procedures are as follows:
\begin{enumerate}
    \item \textit{Matrix structural analysis.} For prescribed updaters $\alpha_e$, perform matrix structural analysis on a classical processor to obtain basic unknowns, i.e., nodal displacement vector $\bm{U}$, which is the solution of the static equilibrium equation $\bm{K}(\alpha) \bm{U} = \bm{F}$ in Eq. \eqref{eq14_objective}.
    \item \textit{Encoding.} Encode the updaters $\alpha_e$ in Eq. \eqref{eq13_onoff} and the slack variable $S$ in Eq. \eqref{eq18_slackencod} with binary variables $q_e$ and $q_c$, respectively.  
    \item \textit{Prepare QAOA circuit.} Define the objective function in Eq. \eqref{eq20_obj3} and transform it to the corresponding Hamiltonian $C$ in Eq. \eqref{eq1_comb} for $\tilde{U}_{C}(C, \gamma)$ in the QAOA quantum circuit in Eq. \eqref{eq2_qaoa1}. 
%
    \item \textit{Parameter fine-tuning and measurement.} Employ conventional optimization procedures or heuristic approach (COBYLA, Bayesian optimization, or FLRS) to adjust the variational parameters such that $(\bm{\gamma}^{*}, \bm{\beta}^{*})= \textrm{arg} \max_{\bm{\gamma}, \bm{\beta}}f(C, \ket{\psi_f})$ (Eq. \eqref{eq7_para}). 
    Then, measure the final quantum state to collapse it into a qubit string $z = z_1z_2 \cdots z_n$. 
%
    \item \textit{Structure update.} Decode the updaters $\alpha_e$ from the qubit string and multiply each decoded updater by the corresponding element stiffness matrix of Eq. \eqref{eq11_updak} in the previous iteration step to update the structure. 
%
    \item \textit{Iteration.} Repeat steps 1 to 5 until convergence is achieved, i.e., the specified conditions are met.
\end{enumerate}
The corresponding flowchart is shown in \fref{fig:flowchart}, and \fref{fig:update} schematizes the progression of optimization as iterations are made.

\section{Numerical examples}\label{sec4}
In this section, two numerical examples are presented to demonstrate the capabilities of the proposed framework with QAOA. 
In the first example, the framework is applied to Case 1 of the six-member truss shown in \fref{fig:truss} to compare three fine-tuning approaches for determining the variational parameters, $(\bm{\gamma}^{*}, \bm{\beta}^{*})$, in Eq. \eqref{eq2_qaoa1}.
Two of them are classical optimizers, COBYLA and Bayesian optimization (hereafter referred to as ``B-opt"), and the other is FLRS, which directly introduces a fixed linear ramp schedule for all variational parameter settings. 
As discussed in Section 2, we propose to use FLRS\cite{Montanez2024protocol}, and the purpose of this example is to illustrate its performance. 
In the second example, structural optimization with QAOA using FLRS is carried out on the truss structure in Case 2 to verify the performance of the proposed framework. 
We also propose an improvement to our proposed framework and discuss its time complexity theoretically. 

\begin{figure}[htbp]
    \centering
    \includegraphics[width=0.4\linewidth]{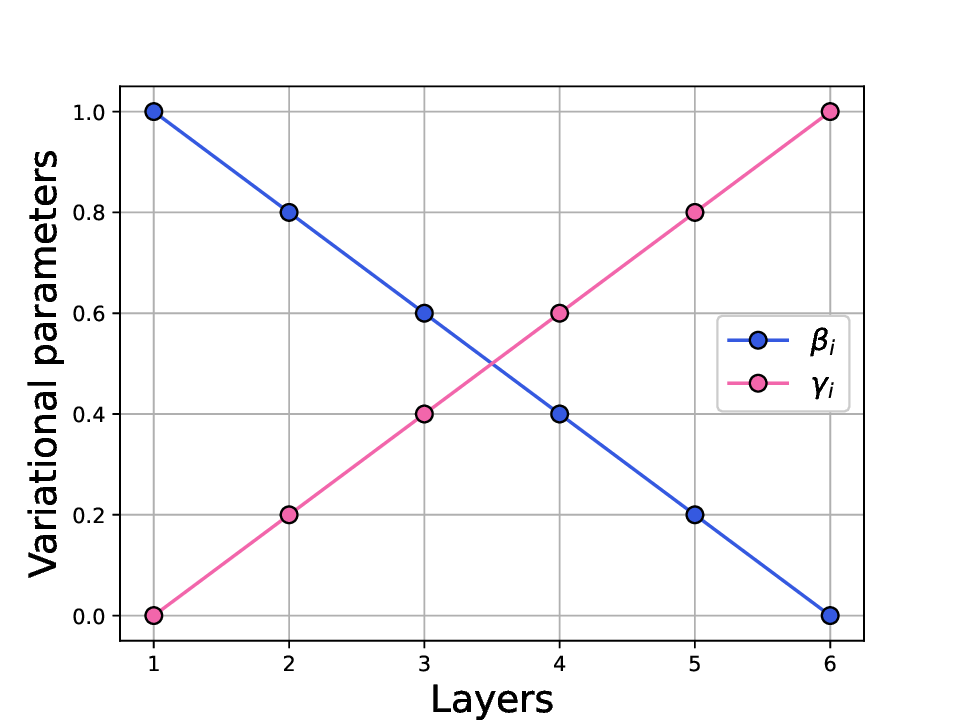}
    \caption{Variational parameter setting for FLRS.
}
\label{fig:parameter}
\end{figure}
\begin{figure}[htbp]
    \centering
    \includegraphics[width=0.5\linewidth]{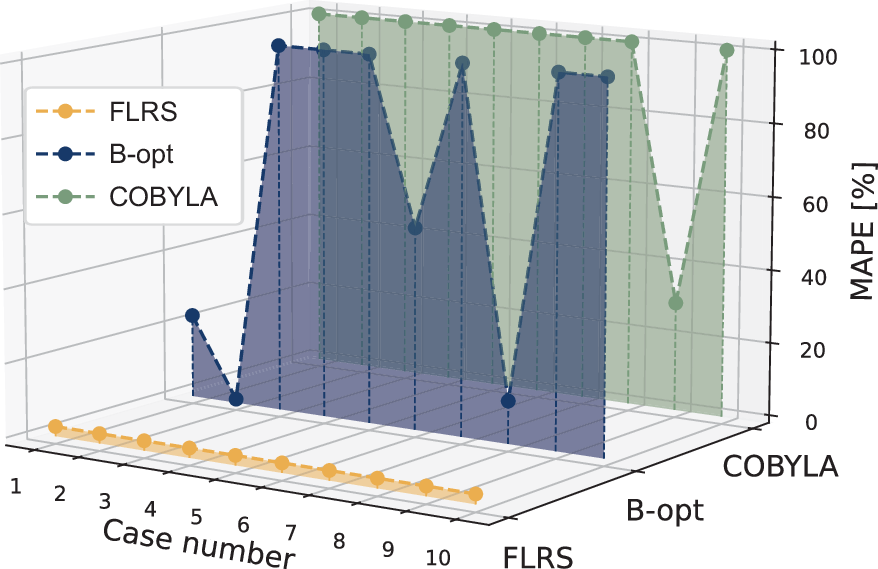}
    \caption{MAPE for 10 executions using three different fine-tuning approaches.
}
\label{fig:optimizer}
\end{figure}

\subsection{A comparative study of parameter fine-tuning approaches for QAOA}
\label{subsec41}

As discussed in \secref{sec2}, among the classical optimizers for QAOA, COBYLA and B-opt are the most popular and widely used to identify appropriate patterns of variational parameters. 
Meanwhile, some studies have also attempted to reduce the computational burden of using classical optimizers when the QAOA contains many layers. 
Among them, this study adopts FLRS \cite{Montanez2024protocol}, which directly introduces a fixed linear ramp schedule for all variational parameter settings.
In the following, these parameter fine-tuning approaches, COBYLA, B-opt, and FLRS, are applied to the $p$-layer QAOA circuit to optimize the configuration of rods for Case 1 in \fref{fig:truss}. 


The number of layers, $p$, is set at 6 in this study, suggesting that there are a total of 12 variational parameters to be optimized. 
Additionally, when defining the objective function in Eq. \eqref{eq20_obj3}, 
a coefficient $\lambda$ must be set as the weight of the volume constraint, and after trying several values in Case 1, $\lambda=5 \times 10^{-1}$ is adopted.
Note that COBYLA and B-opt need to set initial values for the variational parameters to start the optimization process.
The difference is that the former requires a single set of variational parameters, while the latter usually requires multiple sets. 
There is no standard rule for determining initial values in advance for a particular problem.
Random sampling is a commonly used method, but it should be noted that this can lead to uncertainty in the solution obtained.
In this particular example problem, COBYLA is randomly initialized within $[0, 2\pi]$, and the remaining settings are based on default values from \texttt{Scipy} 1.14.1 \cite{2020SciPyNMeth}, which is a popular Python library for implementing COBYLA. 
On the other hand, in this study,
the number of initial observations for B-opt is set to 10, and a total of 100 Bayesian optimization steps are performed. 
According to Kim and Wang \cite{Kim2023to}, 
the the following Matern kernel\cite{genton2001classes} can be employed for B-opt as
\begin{equation}
   k\left(x_i, x_j\right)=\frac{1}{\Gamma(\nu) 2^{\nu-1}}\left(\frac{\sqrt{2 \nu}}{l} d\left(x_i, x_j\right)\right)^\nu K_\nu\left(\frac{\sqrt{2 \nu}}{l} d\left(x_i, x_j\right)\right){,} \label{eq20_Matern} 
\end{equation}
where $d(\cdot, \cdot)$ is the Euclidean distance, $K_\nu(\cdot)$ is a modified Bessel function, and $\Gamma(\cdot)$ is the gamma function. In the following numerical example, the smoothing parameter $\nu$ is set to 0.5, and Upper Confidence Bound (UCB) \cite{Peter2003ucb} is adopted as an acquisition function.
B-opt with such a specification is implemented into our in-house code based on \texttt{scikit-optimize} 0.8.1\cite{sciopt081}. Here, \texttt{scikit} is a popular machine learning Python library that is used as a subroutine in our code. 

Fixed linear ramp schedule (FLRS), on the other hand, assumes a pattern of variational parameters in advance.
According to Montanez-Barrer and Michielsen\cite{Montanez2024protocol}, the variational parameters in the $p$-layer QAOA circuit can be linearly defaulted as 
\begin{equation}
    \beta_i = \left(1-\frac{i}{p}\right)\Delta\beta \quad \text{and} \quad \gamma_i = \frac{i+1}{p}\Delta\gamma,
    \quad (i = 0, \ldots, p), 
\label{eq21_vari}
\end{equation}
where $\Delta\beta$ and $\Delta\gamma$ represent scaling factors and are supposed to be defined individually according to the nature of the particular problem. In this study, we set both two at 1.0. 
The pattern for $p=6$ is shown in \fref{fig:parameter}.

In order to determine the performance of the above three different fine-tuning approaches, an appropriate indicator needs to be defined. 
First, we will perform truss optimization based on the OC method on a classical processor, and the result is called the reference optimal solution and denoted by $y$. Preliminarily, $y \approx 0.13$ was calculated. 
Next, 10 independent optimization runs are performed by QAOA using each approach, and the result is generically denoted by $f(\alpha)$ where $\alpha$ is the approach identifier. 
Then, mean absolute percentage error (MAPE) is used as a measure to evaluate the discrepancy between $y$ and $f(\alpha)$ as
\begin{equation}
    MAPE\; [\%]= \left|\frac{f(\alpha)-y}{y}\right| \times 100{.} \label{eq22_mape}
\end{equation}

The results of 10 independent runs (a total of 30 execution cases) are shown in \fref{fig:optimizer}. 
In this figure, MAPE is forced to be set to 100 if it exceeds 100 so that all the results can be displayed in the same vertical range.
It should be noted that the FLRS-based method sets default values for all variational parameters in advance, so its uncertainty arises only from the QAOA circuit measurements.
Since a sufficient number of measurements ($1e5$) have been set up, the uncertainty due to measurements is considered insignificant for FLRS. In contrast, the other two optimization methods have, in addition to the uncertainty due to measurements, uncertainty due to different initial parameter sets. 
As can be seen in the figure, the B-opt-based and COBYLA-based QAOA show disappointing performance, struggling to produce acceptable results in 10 independent executions. 
In contrast, the FLRS-based QAOA exhibits a low level of error, which resulted from the process of encoding the updater with binary variables. 
From the above comparative results, it is concluded that the FLRS method has proven to be effective, especially when dealing with many variational parameters.
In addition, because FLRS sets the variational parameters directly, it effectively bypasses the optimization process of other classical optimizers, resulting in a much lower time cost than COBYLA or B-opt.

\begin{figure}[htbp]
    \centering
    \includegraphics[width=0.7\linewidth]{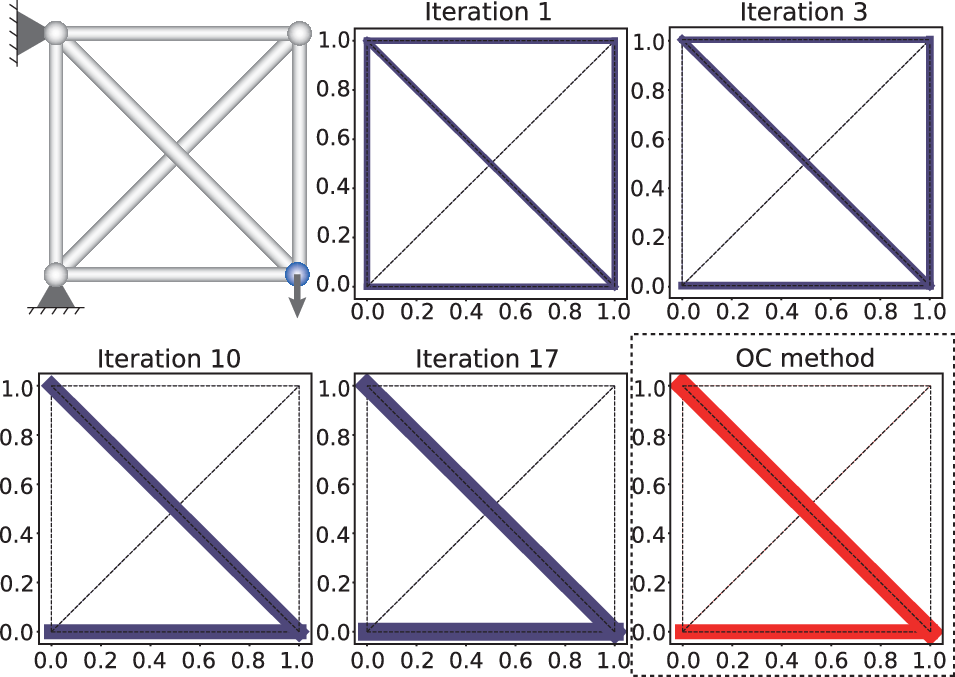}
    \caption{Iterative updating process in Case 1 with FLRS-based QAOA.
}
\label{fig:truss11}
\end{figure}
\begin{figure}[htbp]
    \centering
    \includegraphics[width=0.7\linewidth]{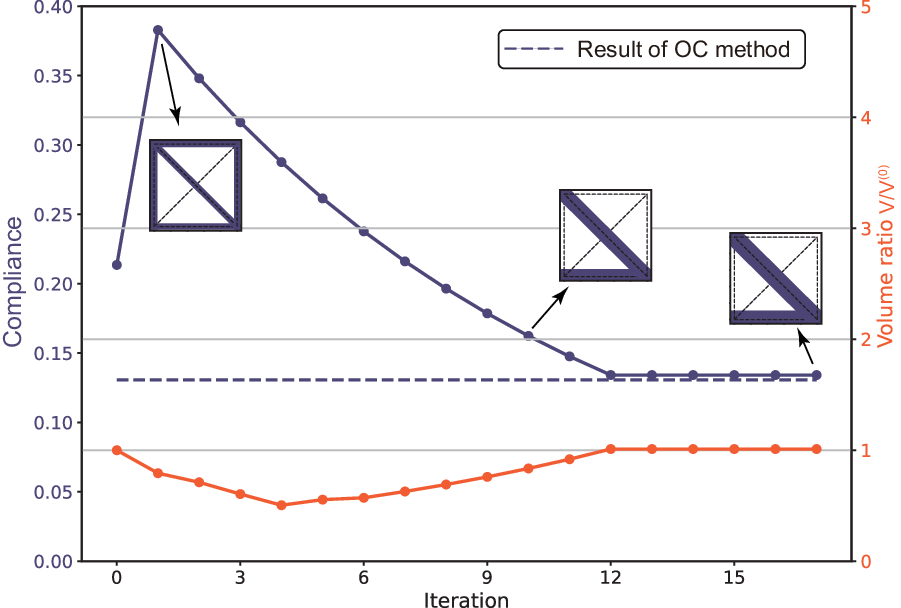}
    \caption{Convergence result in Case 1 wtih FLRS-based QAOA.
}
\label{fig:11process}
\end{figure}

In the following, we will close this section by showing the optimization result of the FLRS-based QAOA.
\fref{fig:truss11} illustrates the specific update process for a six-member truss structure. 
It can be seen that the proposed approach effectively discriminates between critical and non-critical rods and optimizes the distribution of truss members accordingly. 
\fref{fig:11process} shows the convergence trend of the objective function, compliance, and the ratio of total volume to initial volume.
As mentioned earlier, each updater is represented only by two qubits, so four options of $\{\epsilon, 0.1, 1.0, 1.1\} \; (\epsilon \in (1 \times 10^{-10}, 2 \times 10^{-10}))$ are provided.
It is worth noting that when the updater is optimized to $epsilon$, the corresponding rod is considered deleted, but a little volume is left for numerical stability.
After convergence, the volume of the structure slightly exceeds the volume constraint, partly because of this, but also due to encoding, which does not negate the proposed framework. 

\begin{figure}[htbp]
    \centering
    \includegraphics[width=0.7\linewidth]{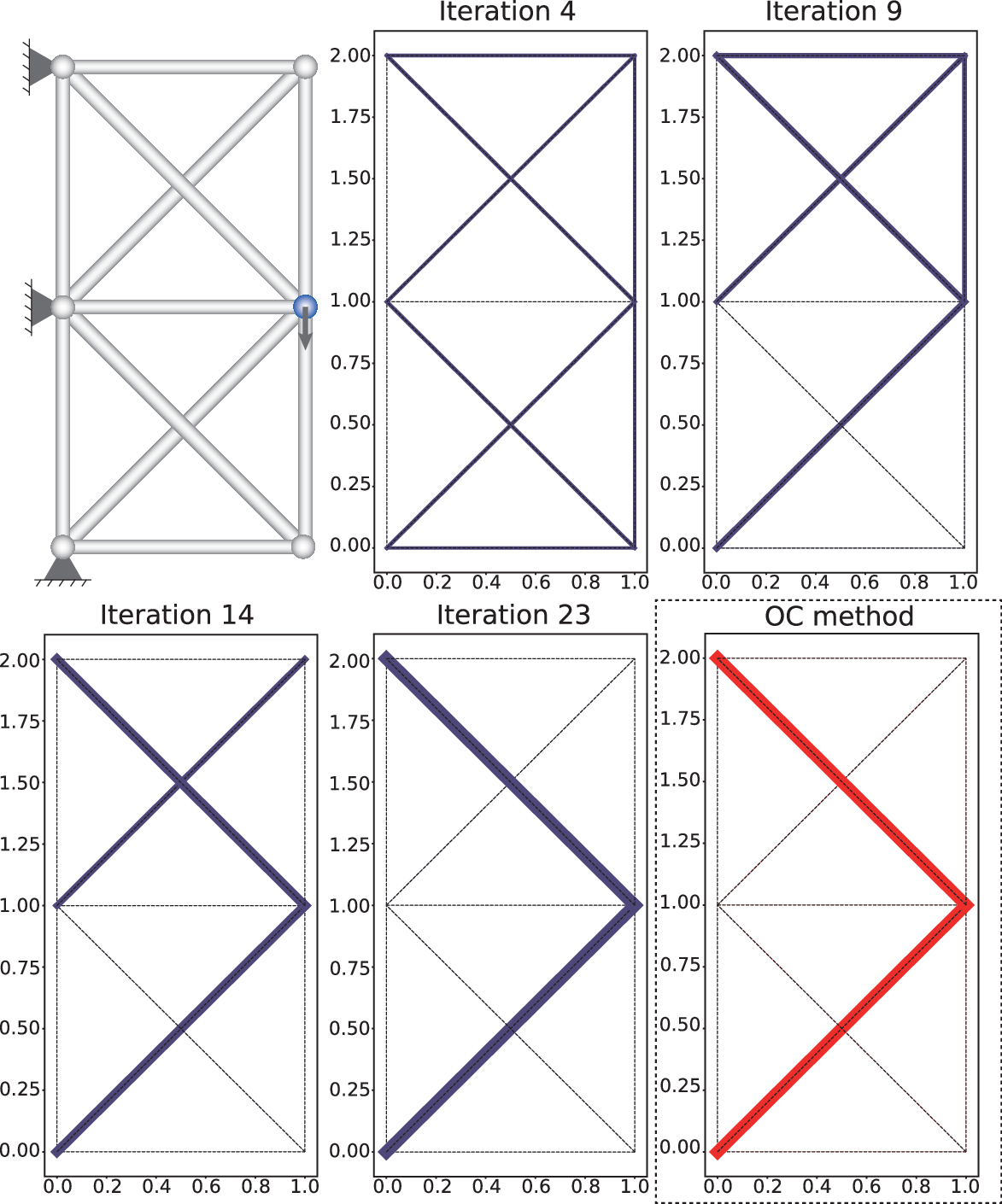}
    \caption{Iterative updating process in Case 2 with FLRS-based QAOA.
}
\label{fig:truss22}
\end{figure}
\begin{figure}[htbp]
    \centering
    \includegraphics[width=0.75\linewidth]{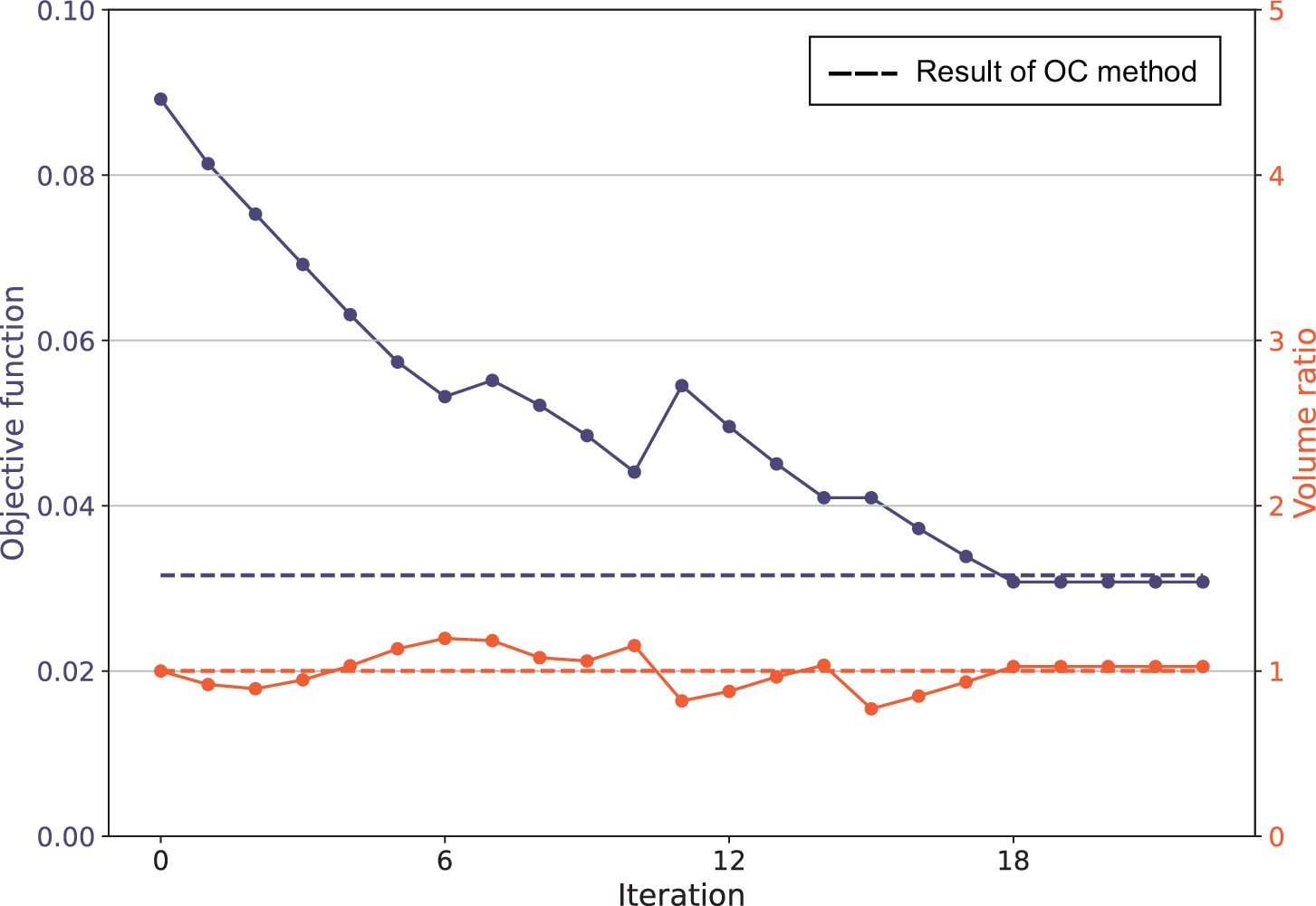}
    \caption{Convergence result in Case 2 with FLRS-based QAOA.
}
\label{fig:22process}
\end{figure}

\subsection{Truss optimization for Case 2}\label{subsec42}

The discussion in the previous section validated the effectiveness of the FLRS method.
Therefore, in this section, only the FLRS-based QAOA is applied to Case 2 of the 11-member truss in \fref{fig:truss}.
Since Case2 contains 11 rods, a total of $24 \; (=11 \times 2 + 2)$ qubits would be employed according to the definition of the objective function in this framework.
The depth (number of layers) of the circuit is set at 8 in this case, and the weight coefficient $\lambda$ that controls the volume constraint is set to $4.25 \times 10^{-2}$, which has been determined by trial-and-error.

\fref{fig:truss22} shows the optimization process in the resulting structural update, together with the result from the OC method on a classical processor.
As can be seen from this figure, the FLRS-based QAOA obtains the converged structure that is consistent with that of the OC method. 
\fref{fig:22process} shows the convergence trend of the objective function, compliance, and the ratio of total volume to initial volume. 
It can be observed that the volume ratio $V/V^{(0)}$ when using the FLRS-based OAOA shows a slight oscillation around the volume constraint of 1.
This is due to the limited number of qubits currently available to encode the updaters and slack variable.
In fact, when the structure volume approaches $V^{(0)}$, QAOA faces a trade-off. That is, it must choose between slightly violating the volume constraint by increasing the cross-sectional area of the critical rod, or removing the critical rod to strictly satisfy the constraint.
Obviously, the former case would be relatively more acceptable.
In the future, as the scalability of quantum hardware, especially the capacity of qubits, improves, such a challenge is expected to be effectively mitigated.
Furthermore, it can be seen from \fref{fig:22process} that the compliance of the truss structure optimized by the FLRS-based QAOA is slightly lower than the OC method.
Specifically, the minimum compliance obtained by the OC method was $3.16 \times 10^{-2}$. 
This is because the converged volume exceeds the volume constraint by about $2\%$, for the same reason as the result in the previous example problem.

Finally, we would like to introduce a strategy for reducing the total cost of the iterative process in QAOA.
As mentioned previously, when using two qubits, on-off encoding provides four options for each updater.

If the cross-sectional area of a particular rod becomes smaller than a negligible value during the iterative update of the QAOA, e.g. less than 1\% of the initial area $A^{(0)}$, there is no need to continue updating that area.
For example, if the corresponding updater is optimized to be $\{0,0\}$ in any iteration, the cross-sectional area of this rod is multiplied by $\epsilon$.
Similarly, if the updater has never been set to $\{0,0\}$ but is optimized to be $\{0,1\}$ multiple times in succession,  the cross-sectional area of this rod decreases by a factor of 0.1 many times to less than $0.01\times A^{(0)}$.

In such cases, stopping the further optimization of these rods can effectively reduce the total number of qubits required for subsequent the iterative process.
As is well known, the query complexity of combinatorial optimization problems can be reduced exponentially by minimizing the number of qubits. 
This strategy, called ``early stopping update'', not only improves the computational efficiency of the proposed framework, but also improves its scalability and competitiveness for dealing with large-scale topology optimization problems in future research.

\subsection{Discussion on time complexity }\label{subsec43}
Given the scale and noise level of current quantum processors, QAOA is still far from demonstrating quantum supremacy. 
Therefore, it is expected that for some period of time in the foreseeable future, applications to structural optimization at a certain scale will have to rely primarily on quantum simulators.
According to Crooks\cite{crooks2018performance}, the time complexity of QAOA applied to the Max-cut problem is $\textit{O}(Np)$ where $N$ and $p$ denote the number of qubits and the number of QAOA layers, respectively. 
Since the form of the objective function in this study is the same quadratic model as the Max-cut problem, this subsection will provide a comparative study on time complexity among
the frameworks of Kim and Wang\cite{Kim2023to}, Xu et al.\cite{Xu2024Gate}, and the proposed framework based on this estimate. 
For the sake of simplicity, we refer to the first two frameworks as K-W and Xu, respectively.

In K-W, each design variable is defined as a binary variable concerning
the inclusion or exclusion of the corresponding rod in the truss. 
Therefore, their method is non-iterative. 
The objective function is then constructed in the form of a ground structure optimization problem as
\begin{equation}
    \begin{aligned}
        f(q_1, \ldots, q_N) = & \text{ } (1-q_1)(1-q_2) \cdots (1-q_N)\delta(\bm{Q}_1)\\
        &+ q_1 (1-q_2) \cdots (1-q_N)\delta(\bm{Q}_2)\\ 
        &+ \cdots + q_1 q_2 \cdots q_N\delta(\bm{Q}_N){,}
    \label{eq25_kim3}
    \end{aligned}
\end{equation}
where $N$ is the number of qubits, which equals to the number of rods, and $\delta(\bm{Q}_i)$ is the displacement at the target node obtained by solving a linear system $\bm{KU} =\bm{F}$ in matrix structural analysis on a classical processor.  
Here, $\bm{Q}_i$ is a binary string consisting of $(q_1, \ldots, q_N)$ to represent a specific situation $i$. 
As can be seen from this equation, a classical processor is forced to face the tedious calculation of polynomial expansion, and there are $2^N$ items in total. 
The complexity of expansion varies from term to term. For example, the time complexity is 2 for $(1-q_1)q_2 \cdot q_N$ and $2^N$ for $(1-q_1)(1-q_2)\cdots(1-q_N)$.
Hence, the time complexity for the expansion becomes $\sum_{i=1}^{N}{^N\mathbb{C}_{i}2^{i}}$ where $^N\mathbb{C}_{i}$ are combinations with $N$ items taken $i$. 
Assuming that the B-opt is adopted as a classical optimizer to optimize the variational parameters and that the QAOA circuit needs to be accessed $k$ times, the total time complexity of K-W is $\textit{O}(pkN+\sum_{i=1}^{N}{^N\mathbb{C}_{i}2^{i}})$.

In Xu, the rod cross-sectional area is represented as a discrete variable using one-hot encoding, and an objective function is defined to minimize the volume of material used to meet the design requirements for load and deformation. 
Additionally, their framework is also non-iterative and requires $h \geq 4$ to ensure the representability of the encoding when using $h$ qubits.
Furthermore, a gradient-based classical optimizer, Adam \cite{Adam2014}, is employed to determine variational parameters. 
Therefore, if we let $k$ be the number of iterations in the optimization calculation using the classical processor, then Xu's time complexity is estimated to be $\textit{O}(pkhN)$.

In the proposed framework, the structure is iteratively updated through updaters acting on the rods, and each updater and the slack variable for volume constraints are encoded by 2 qubits. 
By adopting FLRS, the framework avoids classical optimization.
In addition to this, the major difference from K-W and Xu is that our framework is an iterative design strategy, which requires $I_{\rm d}$ times access to the QAOA circuit
to achieve convergence. 
Furthermore, thanks to our early-stop updating strategy, the consumption of qubits in $I_{\rm d}$ iterations is constantly decreasing.
Denoting $n_{\Delta}$ by the number of rods that are no longer updated in a given iteration by $n_{\Delta}$, 
the time complexity of our framework is estimated as follows:
\begin{equation}
    O(pI_{\rm d} \cdot \sum_{n_{\Delta} \in \Omega} {2N+2-2n_{\Delta}}) \approx O(\zeta pI_{\rm d}N), 
    \label{eq28_expl}
\end{equation}
where $\zeta$ is a coefficient that satisfies the inequality $(1<\zeta<2)$ determined from $N$ and $n_{\Delta}$.

\begin{table*}[]
\renewcommand{\arraystretch}{1.5}
	\centering
	\caption{Comparison of the three frameworks in terms of qubit usage and time complexity.}
	\label{tab:1}  
	\begin{tabular}{cccc ccc}
		\hline\noalign{\smallskip}	
		Frameworks &Qubit usage&Time complexity&Time complexity with FLRS   \\
		\noalign{\smallskip}\hline\hline\noalign{\smallskip}
		K-W framework& $N$ & $\textit{O}(pkN+\sum_{i=1}^{N}{_N\mathbb{C}_{i}2^{i}}$) & $\textit{O}(pN+\sum_{i=1}^{N}{_N\mathbb{C}_{i}2^{i}}$) \\
		Xu's framework& $hN$ & $\textit{O}(pkhN)$&$\textit{O}(phN)$ \\
		Our framework& $2N$ & $\textit{O}(\zeta pI_{\rm d}N)$&-\\
		\noalign{\smallskip}\hline
	\end{tabular}
\end{table*}

\tref{tab:1} provides the information on qubit usage and time complexity of the three frameworks. 
In addition, given that FLRS can be easily implemented in the other two frameworks, we have also added a column showing the time complexity when FLRS is also used in that case.
As can be seen from this comparison table, K-W shows the lowest qubit usage, but has the highest time complexity because the polynomial expansions (Eq. \ref{eq25_kim3}), is computed by a classical processor.
Using FLRS in Xu, the time complexity is not much different from that of our framework, but it requires several times as many qubits as ours.


From the above discussion, it can be concluded that the proposed framework is the most robust choice, balancing flexibility, accuracy, qubit usage, and time complexity.

\section{Conclusion}\label{sec5}
This study proposes an iterative structural optimization framework based on QAOA, which, to the best of our knowledge, is the first attempt to integrate FLRS into QAOA to achieve truss structure optimization.
In particular, this framework overcomes the barren plateau problem caused by multiple variational parameters and improves accuracy and stability due to the FLRS configuration set in this study.
As a key highlight, our framework addresses the gap in enhancing computational efficiency when applying QAOA to structural optimization, and provides a robust option that balances flexibility, accuracy, qubit usage, and time complexity. 
This is demonstrated through numerical examples that include a relatively larger number of rods compared to existing studies. 
In fact, in the numerical example, we succeeded in obtaining the optimal truss structure by using 14 and 24 qubits, respectively, in the 2 truss structures with 6 and 11 rods.
This performance is due to the incorporation of FLRS into QAOA, and and suggests the applicability of the proposed framework to practical structural optimization. It should also be noted that since FLRS is a heuristic method, the variational parameters setting adopted in this study is not necessarily applicable to other structural optimization cases. However, as suggested by He et al.\cite{he2024parameterset}, the assumed setting can be used as an initial parameter combination, and a classical optimizer can be further applied to obtain a more accurate solution for a specific optimization case. Compared with directly applying a classical optimizer, we believe that this strategy is more robust and efficient.

The limitation of the framework is that the weight coefficient $\lambda$ that controls the volume constraint is assumed to be adjusted manually. If this weight coefficient is set too small, 
it will be easy to violate the constraint, and the volume of the optimized structure will be larger than 
the original volume $V^{(0)}$. 
Conversely, if it is too large, QAOA will focus more on satisfying the constraint than minimizing compliance, which will result in the accidental removal of important rods. 
The search for an efficient method of adjusting the value of $\lambda$ is a subject of significance for future research. 
\backmatter

%
\bmhead{Acknowledgements}
This work was supported by JST SPRING (Grant Number: JPMJSP2114).
\vspace{5mm}
\bmhead{On the occasion of the 10th anniversary of the journal's publication}
{\it As someone who has been part of the editorial team since the journal's launch, we are very happy that our paper will contribute to the special issue “Reflecting and anticipating the future of computational mechanics: celebrating the 10th anniversary of AMSES and honoring Prof. Ladev{\`e}ze”. I wish Pierre good health and further development for this journal, which is now led by Paco.} \\

\section*{Declarations}
\begin{itemize}
\item Funding
\\This work was supported by 
JST SPRING, Grant Number JPMJSP2114. 
\item Competing interests
\\The authors declare that they have no competing interests.
\item Ethics approval and consent to participate
\\Not applicable
\item Consent for publication
\\Not applicable
\item Data availability 
\\The data that support the findings of the study are available from the corresponding author upon reasonable request.
\item Materials availability
\\The materials that support the findings of the study are available from the corresponding author upon reasonable request.
\item Code availability 
\\Not applicable
\item Author contribution
\\J. Xiao: Development of the QAOA framework, validation, investigation, and writing of the original draft; N. Sukulthanasorn: Code implementation for truss analysis, conceptualization of truss optimization, discussion, review, and editing of the first draft; S. Moriguchi and R. Nomura: Investigation and review of the original draft; K. Terada: Funding acquisition, conceptualization, methodology development, supervision, review, and editing.
\end{itemize}







\begin{appendices}





\end{appendices}


\bibliography{sn-bibliography}

\end{document}